\documentclass[preprint2]{aastex}

\usepackage{graphicx}
\usepackage{times}

\begin{document}


\title{A Prototype for the PASS Permanent All Sky Survey}

\author{H. J. Deeg, R. Alonso, J. A.  Belmonte}
\affil {Instituto de Astrof\'\i sica de Canarias, C. Via Lactea S/N, 38200 La Laguna, Spain}
\author{Keith Horne, K. Alsubai, A.C. Cameron}
\affil{School of Physics and Astronomy, University of St. Andrews, KY169SS, Scotland, UK}
\author{L.R.  Doyle}
\affil{SETI Institute,515 N. Wishman Ave, Mountain View, CA 94043, USA}


\begin{abstract}
A prototype system for the Permanent All Sky Survey (PASS) project is presented. PASS is a continuous photometric survey of the entire celestial sphere with a high temporal resolution. Its major objectives are the detection of \emph{all} giant-planet tran
sits (with periods up to some weeks) across stars up to mag 10.5,  and to deliver continuously photometry that is useful for the study of \emph{any} variable stars. The prototype is based on CCD cameras with short focal length optics on a fixed mount. A s
mall dome to house it at Teide Observatory, Tenerife, is currently being constructed. A placement at the antarctic Dome C is also being considered. The prototype will be used for a feasibility 
study of PASS, to define the best observing strategies, and to perform a detailed characterization of the capabilities and scope of the survey. Afterwards, a first partial sky surveying will be started with it. That first survey may be able to detect tran
siting planets during its first few hundred hours of operation. It will also deliver a data set around which software modules dealing with the various scientific objectives of PASS will be developed.  The PASS project is still in its early phase and teams
 interested in specific scientific objectives, in providing technical expertise, or in participating with own observations are invited to collaborate.
\end{abstract}



\section{PASS - The Permanent All Sky Survey }
The Permanent All Sky Survey is intended to perform a continuous photometric survey of the entire celestial sphere with a high temporal resolution. Its major objectives are the detection of \emph{all} giant-planet transits (with periods up to some weeks) 
across stars up to mag 10.5,  and the acquisition of photometry that is useful for the study of \emph{any} variable stars, reaching somewhat fainter magnitudes (depending on the precision required). About 250\ 000 stars are expected to be surveyed by PASS
, with the goal to obtain their light-curves with the upmost continuity on time-scales of minutes, and to link these curves into the databases that are commonly used by the astronomical community. The basic instrument design consists of several arrays of 
\emph{fixed} CCD cameras with short (about 50mm) focal length, which will simultaneously survey the entire sky that is visible from their observing location. These cameras will continuously take short exposures in which the stars will appear as short trac
es. A more detailed presentation of the objectives and design of PASS is forthcoming (Deeg et al.~\cite{pass1}).
\begin{figure}[t]
   \centering
   \includegraphics[width=8.5cm]{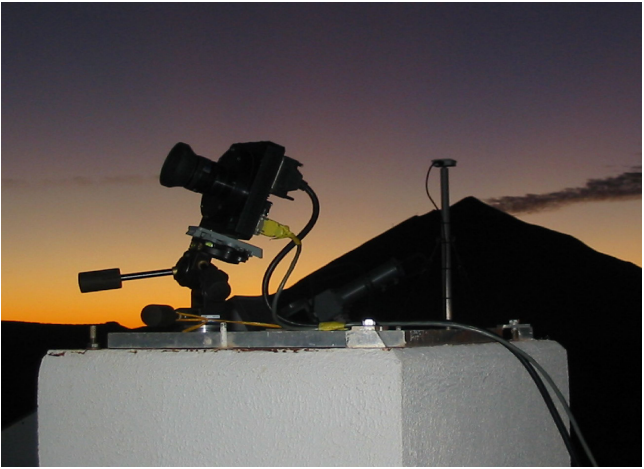}
   \caption{The prototype of PASS at 'first light', 17 January 2004.}
   \label{firstlight}
\end{figure}

\section{PASS0 - A prototype for a feasibility study}
While the basic design of PASS is very simple, it is nevertheless a project that requires a complex and potent data reduction pipeline, and the installation of a considerable number of CCD cameras in several locations. Also, there is little experience in 
the photometric analysis of unguided images, and in the analysis of stellar images taken with optics of very short focal lengths. A study to demonstrate the feasibility of this instrument design is therefore being undertaken. This study started with theor
etical signal-to-noise calculations and the creation of artificial images. A photometric reduction of these images showed that aperture-photometry on trace-shaped stellar images is able to extract light-curves close to the calculated S/N (Deeg et al.~\cite{pass1}). It was also shown, that confusion due to stellar density affects only a small fraction of stars at galactic latitudes of $\approx10 \degr$.
As a next step, a prototype with one CCD camera has been set up on several occasions since January 2004 at Teide Observatory in Tenerife (Fig. \ref{firstlight}).  Funding has been obtained towards the purchase of a further CCD camera, and the construction
 of a small dome is in progress. 
The current prototype consists of two Apogee AP10 cameras with a $2k \times 2k$ front-illuminated Th7899 CCD detector, and Nikon Nikkor F-series lenses with $f=$50mm and maximum apertures of  f/1,2, giving each a field of view of $30\degr \times 30\degr$.
 The cameras are mounted on a heavy-duty tripod-heads (Manfrotto 229) with scales for altitude, azimuth and rotation. A small GPS receiver with USB port (Laipac G10U) will synchronize the control computer in time. Observations with the prototype are aimed
 at these objectives, in rough temporal order:
\begin{itemize}
\item Evaluate if main objectives can be achieved in a  'baseline' situation (good observing site, low to mid airmasses, no moon, average stellar crowding, mid-latitude declination).
\item Define a set of optimum instrumental and operational  parameters (exposure time, best lens-apertures or 'f-stop', possibly evaluating different optics and CCD cameras). 
\item Characterize capabilities and scope of PASS ('detection space') in baseline situation.
\item Characterize capabilities in different situations, from observations under different conditions.
\item Starting a first survey in a limited field (see Section 4.).
\end{itemize}

\begin{figure}[t]
   \centering
   \includegraphics[width=8.5cm]{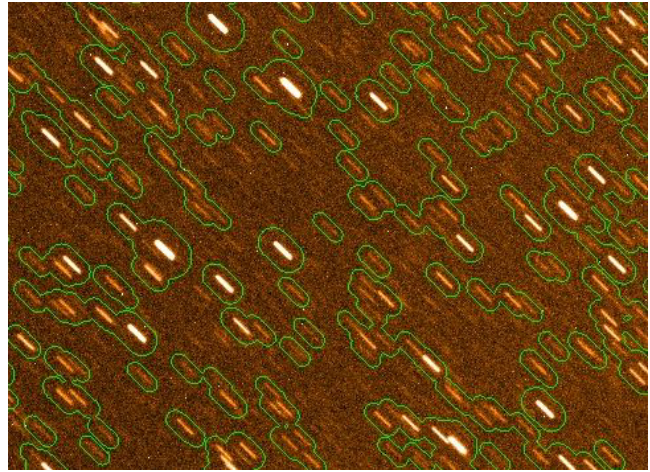}
   \caption{Trace-shaped apertures on which photometry is performed. They are shown over an image from the prototype with 60 sec exposure time. Shorter exposure times, about 15-20 seconds, are expected to constitute the 'baseline' in future observations.}

   \label{apertures}
\end{figure}

While first sequences of images have been taken with the prototype in temporary set-ups, a fixed set-up is required for any observations that will indicate the precision that may be achieved with PASS. In the observations foreseen, the camera will be set 
up in a semi-permanent fashion, and be pointed into the same direction for weeks without disturbance. Such stability is a fundamental requirement for the success of PASS, as unguided images will be taken every night at the same set of sidereal times (ST).
 Thus, in images taken at different nights, but at the same ST, the stars will be moving over exactly the same CCD pixels. This feature will allow the cancellation of any flatfielding errors, be they from the optics or from the CCD. Light-curves will then
 be derived from a comparison of a star against its neighbors, and across sequences of images from many nights taken at identical ST's.
A photometric reduction based partially on modules used in the SuperWASP (Street et al. \cite{swasp03}) experiment is currently under development. This pipeline performs aperture photometry within trace-shaped apertures such as shown in Fig.~\ref{aperture
s}. Any valid evaluation of the precision that can be obtained by the prototype requires however images that were taken in several nights at the same sidereal time, which is pending on the completion of a dome for PASS. 

\section{Dome C as a potential placement}
For a southern station of PASS, a placement at the antarctic Dome C may be an appealing alternative over a site at mid-southern latitudes (e.g. Chile, Australia). The major virtues of that site result from the extremely low seeing values that have been re
ported. One may expect that scintillation noise will be at least 3-4 times lower than from a very good normal observing site. A further advantage comes from the long observing coverage that can be obtained in a single season. Assuming that useful data can
 be obtained for stars with altitudes as low as 15$\degr$ above the horizon, all stars south of -30$\degr$ will be circumpolar at Dome C (-75$\degr$S) and can be observed permanently. For them, observations in a single winter-season at Dome C will corresp
ond to 3-4 s of observations at a mid-latitude site (where they aren't circumpolar). The major technical difficulties are of course caused by the extreme cold, as temperatures below -80C in winter have been reported. The small size and simple structure o
f PASS may however allow a relatively easy and economic adaptation. The only moving part  -- the shutter of the CCD cameras -- would become unnecessary with the employment of CCD detectors that use interline technology. These are similar to those used in 
video-cameras, and large format grey-scale chips up to 24 x 36mm size are being offered with the 'KAI' series by Kodak. Their disadvantage is a lower quantum efficiency. Astronomical cameras with such detectors are currently arriving on the market, and th
e purchase of one is being evaluated. An alternative would be the enclosure of the prototype in an isolated box with an optical window, which is heated to more moderate temperatures around -20C. 
\begin{figure}[t]
   \centering
   \includegraphics[width=8.5cm, bb= 100 0 656 540]{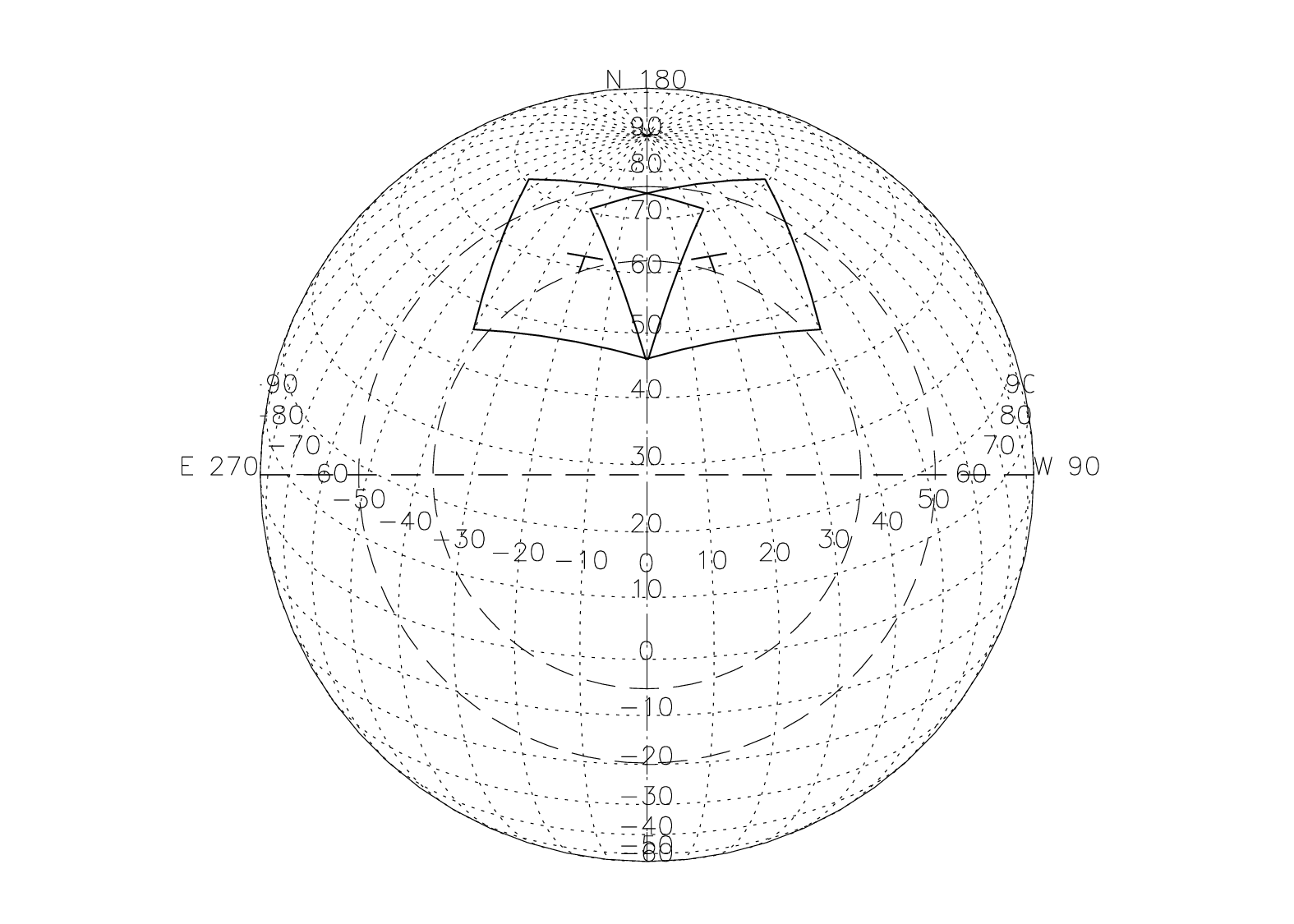}
   \caption{The field of view of the two-camera prototype system for a partial sky-surveying.  It is drawn into an orthographical projection of the sky with celestial coordinates for declination and hour angle, for a location at 28.5$\degr$N (Tenerife). T
he two overlapping squares indicate the $28\degr \times 28\degr$ views of the cameras, which range from declinations of 45$\degr$ to over 70$\degr$. The two dashed concentric rings indicate airmasses of 1.2 (inner one) and 1.5 (outer one).}
   \label{field}
\end{figure}
\begin{figure}[t]
   \centering
   \includegraphics[width=8.5cm]{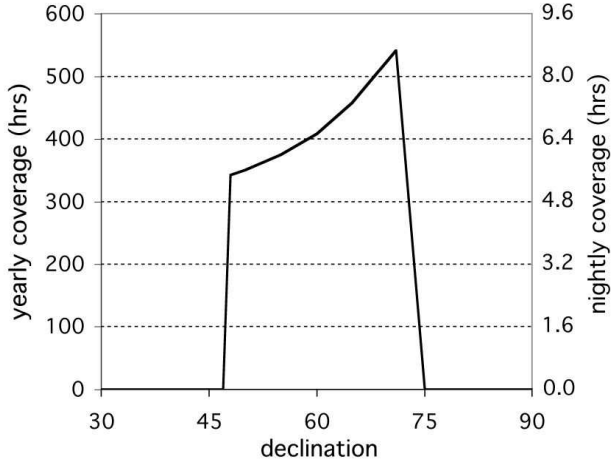}
   \caption{Yearly and nightly coverage that is obtained by the 2-camera prototype, in dependence of declination. Nightly coverage is length of time a star need to pass across the field of view. Yearly coverage gives the hours that star at a given declina
tion can be observed during one year. Variations tied to right ascension, which arise from varying lengths of nights in summer and winter, are ignored here.}
   \label{coverage}
\end{figure}

\section{PASS1 - A first stage of the PASS survey}
After the termination of prototype observations, mainly aimed at evaluating the instrument under varying conditions, the start of the PASS survey in a limited field is planned. This should lead to the first scientifically useful data, and allows to gain o
bservational experience. Also, some parameters affecting performance may only be evaluated from data that are obtained over a longer term, such as dependencies on seeing, temperature, and the effects caused by precession or nutation. That data set will al
so be important in developing and testing the reduction pipeline and calibration procedures. Furthermore, it will be the starting point in the development of procedures for the scientific interpretation of the data, where the development of modules for ea
ch scientific objective is foreseen.  The pointing that is foreseen for a 2-camera system is towards the celestial meridian at mid-northern declinations (Fig.~\ref{field}). This pointing is a compromise between moderate airmasses (best at zenith) and a go
od nightly coverage of stars (best near celestial pole), adapted to a mid-latitude site like Tenerife (28.5$\degr$N). The yearly coverage of a star at a given declination may be derived from the equation:
\[\ \ \ \ \mathrm{T_{cov} = (HA_{max}-HA_{min})/24 \cdot T_{night} }\ ,\]
were $\mathrm{T_{cov}}$ is the yearly observational coverage,  $\mathrm{HA_{max}}$ and $\mathrm{HA_{min}}$ delimit the hour-angle that is coverered (by a fixed instrument), and $\mathrm{T_{night}}$ is the total yearly night time. Assuming a conservative t
otal of 1500 hrs of useful night-hours per year, $\mathrm{T_{cov}}$ may be expected to be in the range of 300-500 hours (Fig.~\ref{coverage}), whereas 300 - 400 hours of observations may be considered a minimum to detect short-periodic planetary transit c
andidates with some reliability (Deeg et al.~\cite{pass1}). The prototype may therefore be expected to find first transits after 1-2 seasons of observations. When a reduction pipeline has been put in place, further cameras may be added at this or other si
tes at any later date, allowing thereby a gradual start of the Permanent All Sky Survey.

\section{Participating in PASS}
It should be noted that the PASS project is still in its early phase and we expect that it provides the opportunity to cover a wide range of topics and objectives. Its initiators welcome any expression of interest that may lead to participation in this pr
oject. We also want to stress, that the modular and extensible nature of PASS, with several camera arrays in various locations, will allow a relatively simple adding of observational equipment by collaborating teams.



\acknowledgements This project is partially being funded by grant AyA 2002-04566 of the spanish national science plan. We also thank the SuperWASP team for the loan of an AP10 CCD camera, which allowed an early start of this project.


\end{document}